\documentclass{aastex}

\begin{document}                                                                                  

\centerline{\bf{The structure and dynamics of young star clusters:}}
\centerline{\bf{ King~16, NGC~1931, NGC~637 and NGC~189}}

\centerline{\bf{Priya Hasan, S N Hasan and Urmi Shah}}
\bf{Department of Astronomy, Osmania University, Hyderabad~-~500007,~India.}


\begin{abstract}
In this paper, using 2MASS photometry, we study the structural and dynamical properties of four young star clusters viz. King~16, NGC~1931, NGC~637 and NGC~189. For the clusters King~16, NGC~1931, NGC~637 and NGC~189, we obtain the limiting radii of $7'$, $12'$, $6'$ and $5'$ which correspond to linear radii of 3.6~pc, 8.85~pc, 3.96~pc and 2.8~pc respectively. The reddening values $E(B-V)$ obtained for the clusters are 0.85, 0.65--0.85, 0.6 and 0.53 and their true distances are 1786~pc, 3062~pc, 2270~pc and 912~pc respectively. Ages of the clusters are 6~Myr, 4~Myr, 4~Myr and 10~Myr respectively. We compare their structures, luminosity functions and mass functions ($\phi(M) = dN/dM \propto M^{-(1+\chi)}$) to the parameter $\tau = t_{age}/t_{relax}$ to study the star formation process and the dynamical evolution of these clusters. We find that, for our sample, mass seggregation is observed in clusters or their cores only when the ages of the clusters are comparable to their relaxation times ($\tau \geq 1$). These results suggest mass seggregation due to dynamical effects. The values of $\chi$,  which characterise the overall mass functions for the clusters are 0.96 $\pm$ 0.11, 1.16 $\pm$ 0.18, 0.55 $\pm$ 0.14 and 0.66 $\pm$ 0.31 respectively. The change in $\chi$ as a function of radius is a good indicator of the dynamical state of clusters.

\bf{star clusters: young -- near-infrared photometry -- 
       color--magnitude diagrams -- pre-mainsequence stars -- initial mass function--relaxation time--
       2MASS}

\section{Introduction}
Star clusters are the most fundamental stellar sytems in understanding the star formation process, which is still full of mysteries. A good study of these stellar systems is the basis of understanding galaxies which are the larger building blocks of the universe \citep{lynga82,janes81,friel95,bonatto05,piskunov06}. 

\end{abstract}
Homogeneous samples of photometric data, coupled with uniform methods of data analysis are essential to make statistical inferences based on the fundamental parameters of clusters. These studies can contribute to understanding the galactic disk, formation and evolution of clusters, molecular cloud fragmentation, star formation and evolution. In this work, we study a sample of young clusters viz.  King~16, NGC~1931, NGC~637 and NGC~189  using photometric data from the Two~Micron~All~Sky~Survey (2MASS) \citep{Skrutskie06}.  The 2MASS covers 99.99\% of the sky in the near-infrared $J$~(1.25~$\mu$m), $H$~(1.65~$\mu$m) and $K_{s}$~(2.16~$\mu$m) bands (henceforth $K_{s}$ shall be refered to as $K$). Hence the 2MASS database has the advantages of being homogeneous, all sky (enabling the study of the outer regions of clusters where the low mass stars dominate) and covering near infrared wavelengths where young clusters can be well observed in their dusty environments. \cite{dutra01} discovered 42 objects at infra-red wavelengths using the 2MASS survey. Many papers devoted to the study of clusters using the 2MASS have been presented in the past few years \citep{bica03,bica06,tadross07,kim06} showing the potential of this database.

We use the 2MASS database to study the sample of four young clusters, to investigate the structure and dynamical state of these clusters close to their time of formation. The sample is selected on the basis that all the four clusters have an age of $\approx$10~Myr reported in literature (see Table 1) and have been formed in different environments. We study the structures and dynamical states of our sample of clusters and determine their mass functions (MFs) and degree of mass seggregation in various regions of the clusters. To study the sample, we construct radial density profiles (RDPs), color--magnitude diagrams (CMDs),  color--color diagrams, luminosity functions (LFs) and MFs. Such studies are not possible using heterogenous datasets where unknown biases may be present. 

The initial mass function (IMF) is the distribution of stars of varying masses from the original parent cloud. The universality of the IMF and the influence of environment on star formation is still a matter of debate. As these clusters are very young (age $\leq$ 10 Myr), their MF may be approximated as the IMF. The sample of clusters have been made from differing initial conditions and subject to varied influences of external interactions with the galactic field, thus leading to observable differences, which we explore.  However, from a recent study of \cite{kroupa07}, even in the case of very young clusters, there is a change in the MF due to the dynamics of young clusters which loose a significant fraction of their stars at an early age. 

Mass segregation is the redistribution of stars according to their masses, thus leading to the concentration of high mass stars near the centre and the low mass ones away from the centre. This has been observed in a variety of clusters, both young and old. The variation of the MF of these clusters is determined in different regions of the clusters and their values are compared.  Further, we estimate from the value of $\tau = t_{age}/t_{relax}$, the degree of mass segregation expected due to dynamical effects and compare it with our observations. The relaxation time $t_{relax}$ is a characteristic time in which there is an equipartition of energy and the high mass stars with lesser kinetic energy sink to the core and the low mass stars move to the outer regions of the cluster \citep{binney08}.  The value of $\tau$  indicates whether an excess of high mass stars in the cores of clusters is a result of dynamical evolution or the imprint of the star formation process itself. The parameter $\tau$ has been described as an evolutionary parameter \citep{bonatto05} which indicates the extent to which the cluster has relaxed. It relates to the core and overall MF flattening.  For large values of $\tau$,  the high mass stars sink to the centre and the low mass stars with high velocities move towards the outskirts and hence the MFs of clusters show large-scale mass segregation and low-mass stars evaporation.
We report the presence of gaps in the main sequence associated to physical processes in stars \citep{kjeldsen91}.


The plan of the paper is as follows: Section 2 describes the clusters in our sample and shows the corresponding RDPs and the values obtained for the limiting radii for these clusters. Section 3 describes the method of selecting cluster members  and the corresponding values of fundamental parameters obtained.  LFs and MFs are described in Section 4 and a comparative study of these clusters is in the concluding Section 5. 

\section{Basic Data and Earlier Observations}
 The RGB images of the target clusters using the DPOSS images are shown in Figure \ref{clusplot} and their parameters are given in  Table~\ref{clusterdata} \citep{dias07}. 

\begin{figure}[h] 
\centering
\caption{Cluster areas (a) King~16 ($JOE$ bands) ($13.37' \times 13.35'$) (b)NGC~1931 ($JKN$ bands)($13.37' \times 13.35'$) (c)NGC~637($JKN$ bands) ($13.37' \times 13.35'$) (d)NGC~189 ($JKN$ bands)($11.87' \times 11.87'$)}
\label{clusplot}
\end{figure}

King 16 has been studied by \cite{maciejewski07} using $BV$ photometry and they obtained a reddening value $E(B-V) = 0.89$,  age 10~Myr and distance 1920~pc .  It lies close to the clusters Dias~1 and Berkeley~4.


The young cluster NGC~1931 is situated in the extension of the Perseus arm \citep{pandey86}. The cluster shows variable reddening with $\Delta E(B-V)=0.45$ and is at a distance of 2170~pc and an age 10~Myr \citep{bhatt94}. The nebulous cloud in the central region is in the background as inferred by \cite{pandey86} based on the reddening determinations in different regions of the cluster.


NGC~637 has been studied by \cite{grub75} in the $RGU$ photographic system. Photoelectric observations in the $UBV$ system were made by \cite{huest91} to obtain a distance of 2500~pc, reddening 0.66 and age 15~Myr. \cite{phelps94} also observed this cluster and obtained a younger age of 0--4~Myr. A conspicuous gap was found in its color--magnitude diagram, which is not a result of incompleteness of data. \cite{piet06} presented $VI$ photometry of this cluster and monitored the cluster for variables.

NGC~189 is a young compact cluster in the vicinity of Stock~24 and Do~12. It has been studied by \cite{balazs61} and the distance to this cluster was found to be 790~pc .

   In this paper, we have used the 2MASS database.  The point-source $S/N =10$ limit is acheived at or fainter than $J=15.8^{m}$, $H=15.1^{m} $ and $K =14.3^{m}$ for virtually the entire sky and hence we have used the above criteria  to extract the 2MASS data using Vizier~\footnote{http://vizier.u-strasbg.fr/cgi-bin/VizieR?-source=II/246}. Further, we have also added the constraint that photometric errors in each band are $\leq 0.2^{m}$.

\begin{table*}
\small
\caption{Basic cluster parameters \cite{dias07}\label{clusterdata}}
\begin{tabular}{lllll}
\tableline
Parameter & King 16 & NGC 1931 & NGC 637 & NGC 189\\ 
           \tableline
RA(2000)(h:m:s) &00 43 45&05 31 25 & 01 43 04 &00 39 35\\
Decl.(2000)(d:m) &+64 11 08  & +34 14 42& +64 02 24&+61 05 42\\
Galactic longitude &$122.09^{0}$& $173.89^{0}$  &$128.54^{0}$ &$121.49^{0}$\\
Galactic latitude &$+01.32^{0}$ & $+00.28^{0}$&$+01.73^{0}$ &$-01.74^{0}$\\
Ang.diameter & 17.6$^{'}$&5$^{'}$ & 3$^{'}$&5$^{'}$\\
Distance(pc) &1920 &3086 & 2160&752\\
$E(B-V)$(mag) &0.89 &0.738 &0.634 &0.42\\
log(age) &7.00 &7.002 & 6.980 &7.00\\ \hline
\end{tabular}

\end{table*}

\subsection{Determination of Radial Density Profiles}
\label{clusrad}
						     
For accurate determination of the cluster parameters, it is essential to have 
the knowledge of the radial extent of the clusters. Mass segregation might lead 
to a larger `true' cluster size than stated  in the \cite{dias07} catalogue. As the 2MASS data offers all sky coverage we have the opportunity to study farther reaches of the clusters. 

%
%
%
%
%
%

 The centers of the clusters are determined using a program which, given an eye estimated center and radius, counts the number of stars and calculates the average $\bar{X}$ and $\bar{Y}$ of the stars within the radius. If the difference in the position ($\bar{X}$,$\bar{Y}$) from the eye estimated center is smaller than a given tolerance value (a pixel), then  the eye estimated center is taken as the center. If larger, then ($\bar{X}$,$\bar{Y}$) is taken as the new approximate center. The same procedure is repeated iteratively until the difference in the position ($\bar{X}$,$\bar{Y}$)  and the center lies within the tolerance value \citep{sag98}. An error of a few arc seconds  is expected in locating the center. 

\begin{figure} 
\centering
\caption{Radial density profiles (a) King~16 (b) NGC~1931 (c) NGC~637 (d) NGC~189}
\label{radall}
\end{figure}

For the determination of the radial surface density of stars $\rho(r)$ in a cluster, a number of concentric circles with respect to the estimated center are made in such a way that each annular region contains statistically significant number of stars. The number density of stars, $\rho_{i}$ in the $i^{th}$ region is calculated as $\rho_{i}=N_{i}/A_{i}$, where $N_{i}$ is the number of stars in the $i^{th}$ region of area $A_{i}$. 
The RDPs for the clusters are shown in the Fig.~\ref{radall}. The $\chi ^{2}$ minimization technique was used to fit the RDPs to the function $$\rho(r)=\frac{\rho_{0}}{1+(r/r_{c})^2}$$ \citep{king62} to determine $r_{c}$ and other constants. The cluster's core radius $r_{c}$ is the radial distance at which the value of $\rho(r)$ becomes half of the central density, $\rho_{0}$. Probable members are selected from all the stars in the cluster area which satisfy the photometric criterion \cite{walker65} described in the next section. The best fits are shown in the figures with dotted lines for all the stars in the cluster field and solid lines for probable members.  The reduced $\chi ^2$ (variance of residuals) for the fits to the clusters King~16, NGC~1931, NGC~637 and NGC~189 were 0.95, 1.14, 0.98 and 0.89 respectively. The limiting radius of the cluster is the distance from the centre at which the star density becomes approximately equal to the field star density. 
As is evident from the Fig. \ref{radall}, we obtained a radius of $7'$ and  $6'$ in the case of King~16 and NGC~637. 
In the case of NGC~1931, due to the presence of the obscuring nebula, a large number of cluster stars are hidden and hence the RDP was plotted using field stars and we obtained a size of $12'$. In the case of NGC~189, we obtained a size of $5'$.

The  new sky coordinates of the cluster centers for epoch 2000, core and limiting radii and background densities obtained by fitting to  King's profile are given in Table \ref{raddata}. 
\begin{table*}
\small
\caption{Structural parameters from RDPs \label{raddata}}
\begin{tabular}{lllll}
\tableline
Parameter & King 16 & NGC 1931 & NGC 637 & NGC 189\\ 
           \tableline
RA (h:m:s)&00 43 24  & 05 30 4.8   &  01 43 12      &  00 40 05          \\
Decl (d:m:s)& +64 11 00 &  +34 13 22 & +64 02 19.8   & +61 05 00     \\
Field density &0.97  $\pm$ 0.27    & 1.26 $\pm$ 0.03  & 0.08  $\pm$ 0.04  & 0.42 $\pm$ 0.36\\
Core radius   & 0.89 $\pm$ 0.37    & 0.63 $\pm$ 0.16  & 0.36  $\pm$ 0. 13  &1.38  $\pm$ 0.85  \\    
Limiting radius &  7 $\pm$ 1.2 & 12 $\pm$ 0.5 & 6 $\pm$ 0.2 & 5 $\pm$ 0.6 \\
\hline

\end{tabular}

\end{table*}

\section{Membership, Color--Magnitude and Color--Color Diagrams }
VizieR was used to extract $J$, $H$ and $K$ 2MASS photometry of the stars in a circular area of radius $30'$ from the approximate center obtained in the earlier section. The apparent CMDs for the clusters obtained by extracting stars from the areas equal to the sizes obtained from the RDPs and an offset field of the same area are shown in the Fig. \ref{appall}.  In the case of King~16, we extracted data for the cluster within a radius of $7'$ and an the offset area made up of a concentric ring of radius $29.17'$ to $30'$. For NGC~1931, the cluster size determined from the RDP was $12'$ and the extracted offset area was made up of a concentric ring of radius $27.49'$ to $30'$. In the case of NGC~637, the cluster area was within a radius of $6'$ and the offset area of $29.4'$ to $30'$. In the case of NGC~189, as Stock~24  and Do~12 are very close to the cluster, we marked out a field area at a distance of $7'$ to the south-west of the cluster. 

\begin{figure} 
\centering
\caption{Apparent color--magnitude diagrams for the clusters and an offset field 1(a) King~16; 1(b) Offset Field for King~16 1(c) `Clean' CMD of King~16; 2(a) NGC~1931; 2(b) Offset Field for NGC~1931 (c) `Clean' CMD of NGC~1931; 3(a) NGC~637; 3(b) Offset Field for NGC~637(c) `Clean' CMD of NGC~637; 4(a) NGC~189; 4(b) Offset Field for NGC~189(c) `Clean' CMD of NGC~637}
\label{appall}
\end{figure}

To study the intrinsic cluster CMDs, we use the field star decontamination procedure similar to the one applied by \cite{bon2006a,bica06}. In this method, we divide the CMD into cells and count the number of stars in the field and in the cluster area. Assuming that the number of field stars is constant, we randomly remove in each cell,  candidate field stars equal to the number expected in the field to obtain a `clean' cluster CMD. Considering that the solid area in the offset area is different from that in the cluster area, we multiply the number of stars of the offset field by a constant. Figure \ref{appall} shows the field star decontaminated or `clean' cluster CMDs. In crowded field regions, the field star density at fainter magnitudes may be larger than that of the cluster area, thus artificially truncating the main sequence. As this method artificially removes stars and distorts the RDPs, we used this method only to uncover the cluster CMDs and color--color diagrams. To study the cluster structure, LF and MF we use the probable members obtained by the photometric criterion \citep{walker65}. 

The observed data has been corrected for interstellar reddening using the coefficients given by \cite{dutra02} where $A_{J} = 0.856 \times E(B-V), A_{H} = 0.546 \times E(B-V), A_{K} = 0.366 \times E(B-V), E(J-H) = 0.31 \times E(B-V), E(H-K) = 0.18 \times E(B-V)$ where $E(B-V)$ denotes the color excess for the cluster. The clusters King~16, NGC~189 and NGC~637 show uniform reddening and hence reddening values have been obtained by isochrones fits. 

In the case of King~16, the only spectroscopic data we have is that of BD~6384 which has a reddening value of 0.91 and distance modulus (DM) 11.26. Using isochrone fits to the CMD, we get a reddening of 0.85 and a DM of 11.3, which also agrees with the values obtained by \cite{maciejewski07}. 

In the case of NGC~1931, which shows differential reddening \citep{pandey86,bhatt94}, the entire cluster region was divided into 9 regions for which the reddening values were determined individually by isochrone fits. Stars were then corrected for their reddening values depending on their spatial location. We obtained reddening values $E(B-V)$ ranging from 0.65--0.85 and a distance of 3062~pc to the cluster.

For NGC~637 and NGC~189 isochrone fits to the CMDs gave reddening values of 0.6 and 0.53 and distances of 2270~pc and 912~pc respectively.

\begin{figure}
\centering
\caption{NGC 1931: Differential reddening}
\label{red1931}
\end{figure}


To determine the  membership we use two criteria: the radial extent and the photometric criterion described by \cite{walker65}. The photometric criterion is made by plotting a color--magnitude filter along the isochrone with a width of $\approx$ 0.1 in the $(J-H)$ direction and $\approx$ 1.0 in the $J$ direction. Thus we identify main sequence members which may have have been displaced from the main sequence track either due to photometric errors, effects of binarity, etc. Similar filters have also been made in the $H$ vs $(J-H)$ and $K$ vs $(J-K)$ planes. 
The Walker method is valid only for main sequence stars while other luminosity classes and  groups require 
different methods for member identification. 

The absolute CMDs with the isochrones used to determine their ages are shown in the Fig. \ref{absall}. The clusters show a well-defined main sequence, with even the most massive stars still on the main sequence. 
 
 It is interesting to note that three of the cluster CMDs show a gap for early type stars. A gap in the main sequence is loosely defined as a band, not necessarily perpendicular to the main sequence, with no or very few stars. \cite{mer76} pointed out a gap at $(B-V)_{0}=-0.1$, $(J-H)_{0}=-0.05$ possibly related to the way in which the Balmer jump and the Balmer lines behave in late $B$ or early $A$ stars. In the CMDs of King~16 a gap is noticeable at $M_{J}=-3$, $(J-H)_{0} = -0.1$. NGC~1931 has a gap at $M_{J}=-2.5$, $(J-H)_{0} = -0.1$ and  NGC~637 has a gap at $M_{J}=-1.5$, $(J-H)_{0} = -0.1$ . All these are associated to $OB$ type stars. 

\begin{figure}
\centering
\caption{Absolute CMDs (a)~King~16 (b)~NGC~1931 (c)~NGC~637 (d)~NGC~189}
\label{absall}
\end{figure}

\begin{figure}
\centering
\caption{Two--color diagrams (a) King~16 (b)~NGC~1931   (c)~NGC~637 (d)~NGC~189}
\label{redall}
\end{figure}

The unreddened color--color diagrams $(J-H)_0$ versus $(H-K)_0$ for the field star decontaminated clusters are shown in the Fig.~\ref{redall}, indicating the appropriate reddening correction.  In the case of NGC 1931, the plot clearly shows the differential reddening and the possible classical T Tauri stars in the cluster \citep{lada92}.

Table \ref{allpar} shows the values of the fundamental parameters of reddening, distance and age obtained for the clusters using the isochrones \citep{girardi02} and compares it to that obtained by earlier authors.

\begin{table*}
\small
\caption{Parameters estimated for King~16, NGC~1931, NGC~637 and NGC~189}
\begin{tabular}{lllll}
\tableline
   Cluster  & Reddening & Distance(pc) & Age(Myr)& Reference  \\
             \tableline
  King 16 &0.89 & 1920    &  10     & \cite{maciejewski07}\\
         &0.85  & 1786    &  6     & This work\\ \tableline
 
 NGC 1931 & 0.33-1.2      &2160     &       & \cite{pandey86}\\
          &0.55-1.0 & 2170    & 10      & \cite{bhatt94}\\
          &0.738    &	3086&	10       &\cite{lok2001}\\
  
          & 0.65-0.85 & 3062    &   4    & This work\\ \tableline
	
NGC 637   & 0.63 & 2160    & 9.5   &\cite{grub75}\\
          & 0.66 & 2500    & 15      & \cite{huest91}\\
          & 0.65 & 2884    & 4     & \cite{phelps94}\\
          & 0.55 & 2679-3221  & 4   &\cite{piet06} \\
          & 0.6    & 2270  &    4     & This work\\ \tableline	

 NGC 189 &  & 790    &       &\cite{balazs61} \\
        & 0.53  & 912    & 10      &This work \\  \tableline 
		     	       
\end{tabular}
\label{allpar}
\end{table*}

\section{Luminosity and mass functions}

\begin{figure}
\caption{Luminosity functions (a) King~16 (b) NGC~1931 (c) NGC~637 (d) NGC~189 ($J$ in blue, $H$ in green and $K$ in red)}
\label{lfall}
\end{figure}

The LF obtained for clusters using observations has to be corrected for the following three factors: (i) fraction of cluster area studied (ii) completeness of data (iii) field star contamination. As the 2MASS data has 99.99\% completeness for the magnitude range considered and we have  extracted the complete cluster area, we only had to correct the LF for field star contamination. The LF was found for members based on the photometric criterion \citep{walker65} in the $J$ vs $(J-H)$ plane using color--magnitude filters. These filters are lines parallel to the isochrone track with a width of $\approx$0.1 in the $(J-H)$ direction and $\approx$1.0 in the $J$ direction.  A similar color--magnitude filter was applied for the apparent CMDs of the field area shown in Fig \ref{appall}. Thus, we obtain the approximate number of stars which are probable non-members, but still lie within our color--magnitude filter. The number of field stars in each magnitude bin was then subtracted from the number of stars in the cluster area. The LFs in other bands were found by using color--magnitude filters in the $H$ vs $(J-H)$ and the $K$ vs $(J-K)$ plots to identify probable members. Similar filters were also made for the offset field to correct for field star contamination. Figure \ref{lfall} shows the uncorrected (dotted line) and corrected (solid line) LFs for the four clusters in the $J$, $H$ and $K$ bands.   
 
\begin{figure}[h]
\caption{King 16: Mass function}
\label{mfk16}
\end{figure}

\begin{figure}[h]
\caption{NGC 1931: Mass function}
\label{mf1931}
\end{figure}

\begin{figure}[h]
\caption{NGC 637: Mass function }
\label{mf637}
\end{figure}

\begin{figure}[h]
\caption{NGC 189: Mass function}
\label{mf189}
\end{figure}

The MFs were constructed from the LFs using the isochrones \cite{girardi02} with the appropriate ages and distances and fitting them to a fourth order polynomial to find the mass--luminosity relation. The mass function, $\phi(M) = dN/dM \propto M^{-(1+\chi)}$, is an indicator of the star formation process.  The relaxation times for the core and overall clusters have been calculated using the formula $t_{relax}=\frac{N}{8ln N}\times t_{cross}$ where $t_{cross}= R/\sigma_v$, $N$ is the number of stars, $R$ is the radius and $\sigma_v$ is the velocity dispersion. We have used the value $\sigma_v$= 3 km s$^{-1}$ \citep{binmer}. 

The clusters were divided into regions so as to obtain a significant number of stars in each region. For King~16, we obtained a core radius $0.89 \pm 0.24$. We divided the cluster into three regions: core ($0'-0.89'$), halo1 ($0.89'-3'$) and halo2 ($3'-7')$. In the case of NGC~1931, we divided the cluster into three regions: halo1 ($0'-4'$), halo2 ($4'-8'$) and halo3 ($8'-12'$). Due to the central obscuring nebula, we have no stars in the core region, and hence we excluded the same. In the case of NGC~637, we divided the cluster into three regions: core ($0'-0.4'$), halo1 ($0.4-3$) and halo2 ($3'- 6'$). As there were no stars in the core of NGC~189 and the cluster is very small, we divided the cluster into only two regions  : halo1  ($0-2.5'$) and halo2 ($2.5'-5'$).

  The  values of $\chi$ for different regions of the clusters are also indicative of mass seggregation and are shown in Table \ref{dypar} with the mass estimates for each region. The mass estimates for the clusters King~16, NGC~1931,  NGC 637 and NGC~189 are $1382 \pm 44 M_{\odot}$, $848 \pm 14 M_{\odot}$, $583 \pm 6 M_{\odot}$ and $94 \pm 3 M_{\odot}$ respectively using the observed mass ranges. These are the lower limits of the masses for these clusters.

\begin{table*}
\small
\caption{Dynamical Parameters estimated for King~16, NGC~1931, NGC~637 and NGC~189}
\begin{tabular}{lllll}
\tableline
   Cluster & R (arc min) &  $\chi$  &  N  &  mass($M_{\odot}$ ) \\
             \tableline
  King 16   & & 1.4--17.44 $M_{\odot}$   &       & \\ \tableline
   core     & 0--0.89 & -0.44 $\pm$0.10   &  10     & 16 $\pm$ 1 \\
    halo1  &  0.89--3& 0.95 $\pm$ 0.14   &  129     & 811 $\pm$ 26\\
   halo2  & 3--7 & 0.89 $\pm$ 0.18 &  193     & 811 $\pm$ 26\\          
  overall       &  & 0.96 $\pm$ 0.11    &      & 1382 $\pm$ 44\\ \tableline
 
 NGC 1931  &    &  2.39--14.44 $M_{\odot}$    & &\\ \tableline
   halo1         & 0--4 & 0.029 $\pm$0.13   &   39    & 105 $\pm$ 4\\
        halo2    & 4--8& 1.22 $\pm$ 0.18  &  100     & 265 $\pm$ 12\\
         halo3    & 8--12& 1.32 $\pm$0.19 &  184     & 516 $\pm$ 24\\          
   overall      &  & 1.159 $\pm$0.18   &       & 848 $\pm$ 14\\ \tableline
	
NGC 637 &    &   1.6--17.14 $M_{\odot}$    & &\\ \tableline
   core         &  0--0.4&  -1.00  &  7     & 35 $\pm$ 1\\
   halo1         & 0.4--3 &  0.39 $\pm$0.13  &  101     & 328 $\pm$ 24\\
   halo2          &3--6 &    1.18 $\pm$ 0.13 & 123   & 217  $\pm$ 44\\          
   overall      &  &  0.55 $\pm$0.14  &       &583 $\pm$ 6 \\ \tableline  

 NGC 189 &    &  0.86--4.75  $M_{\odot}$    & &\\ \tableline
   halo1         & 0--2.5 & 0.68 $\pm$0.22   &  33    & 61 $\pm$ 3\\
   halo2         &2.5--5  & 0.087 $\pm$0.02   & 14      &36 $\pm$ 1 \\
     overall       &  & 0.66 $\pm$0.31   &       &94 $\pm$ 3 \\  \tableline 
			     	       
\end{tabular}
\label{dypar}
\end{table*}

As observed in the Fig \ref{mfk16}, for the cluster King 16, the $\chi$ value   was found to be 0.96 $\pm$ 0.11  for the overall cluster, -0.44 $\pm$ 0.10  in the core region, 0.95 $\pm$ 0.14  in halo1 and 0.89 $\pm$ 0.18 in halo2.  (Errors in $\chi$ values in this section are aysymptotic standard errors.) The relaxation time is 2~Myr for the core  and 10.5~Myr for the overall cluster. The age of the cluster based on the isochrone fit is 6~Myr. The mass of the most massive star is 17 $\pm$ 1 M$_{\odot}$, which has a nuclear age, $t_{nuc}=10^{10} \times (\frac{1}{M/M_{\odot}})^{2.5}= 7.9$~Myr, which is the upper limit to the age. Since the age of the cluster is much larger than the relaxation time of the core, the core has dynamically relaxed. However, the halo of the cluster has not yet relaxed, hence we see no change in the $\chi$ values of the inner and outer halos. 

 In the case of NGC~1931, as seen in Fig \ref{mf1931}, the MF has been found in three concentric rings each at a radius increasing by $4'$. The regions have the mass functions 0.02 $\pm$ 0.13, 1.22 $\pm$ 0.18 and 1.32 $\pm$ 0.19 respectively. This indicates a rearrangement of stars as a function of distance and a certain degree of mass seggregation. The relaxation time for NGC~1931 is 20~Myr and the upper limit of its age based on the most massive star on the main sequence is 3.5~Myr and its age based on isochrone fits is 4~Myr. The overall cluster has a mass function of $\chi=$1.1 $\pm$ 0.18. We conclude that the innermost halo1 ($0$--$4'$) has relaxed, while the rest of the cluster is in the process of relaxation.

 In NGC~637, as in Fig \ref{mf637}, the $\chi$ values were 0.66 $\pm$ 0.14, -1.0, 0.39 $\pm$ 0.13 and  1.18 $\pm$ 0.13 for the overall, core, halo1 and halo2 of the  cluster respectively. The cluster age by isochrone fits has been found to be 4~Myr and based on the most massive star (15 $\pm$ 1 M$_{\odot}$) of the cluster  is 11.5~Myr. The relaxation times for the core and the overall cluster  are 0.039~Myr and 7.6~Myr respectively. Hence the core has relaxed and the overall cluster has also partially relaxed as is indicative by the value of $\tau$.

 For the cluster NGC~189, as seen in Fig \ref{mf189}, the age of the cluster is around 10 Myr, which is larger than the $t_{relax}$ of 3.59 years. Hence, we see the overall cluster MF is flatter $\chi$=0.66 $\pm$ 0.31 , while the halo1 and halo2 have $\chi$ values 0.68 $\pm$ 0.22 and 0.09 $\pm$0.02  respectively. Halo2 appears flatter since probably it has already started losing low mass stars to its outer environment.

 \section{Conclusions}
 In this paper we have studied four young clusters of comparable ages to understand their structure and dynamics. The RDPs of the clusters have been plotted and the parameters for the clusters such as reddening, distance and age have been determined using isochrone fits. 

We have also plotted the LFs in the $J$, $H$ and $K$ bands and used the derived mass--luminosity relation to find the MFs using all three bands independently. The $\chi$ values have been determined for different regions and the overall clusters as a function of the parameter $\tau$. We use the difference in $\chi$ values to estimate the level of mass segregation of the clusters and their cores. 

In the case of King~16, where $\tau$ = 3  for the core, the core is clearly relaxed as in indicated by its flat $\chi=-0.44$, while for the outer regions where  $\tau$ = 0.57, the cluster has begun to relax. In the case of NGC~1931, the core is relaxed and the outer region  seems to have begun relaxation. In the case of NGC~637, the core is relaxed ($\tau=4/0.039$), and there  appears to be a redistribution of stars in the cluster, indicated by a progressive increase in $\chi$ from 0.39 to 1.15. The outer halo has a larger value of $\chi$ compared to the overall cluster (0.55) as, it appears that the inner halo has thrown away a large number of low mass stars to the outer halo, thus causing an excess in low mass stars and a steeper $\chi$. In the case of NGC~189, where $\tau$ =2.79, the cluster has already relaxed, as is indicative of the flat $\chi$ of the core. The outer core has a flatter $\chi$, probably because low mass stars which were thrown out of the inner halo during relaxation, have been lost and hence the outer halo has a deficit of low mass stars. 

 As, seen from our analysis, mass seggregation is observable in the cores of the clusters King~16, NGC~1931 and   NGC~637  where the cluster ages are comparable to the relaxation times. In the case of NGC 189, where the relaxation time is lesser than the age of the cluster, we notice a flatter mass function. This  implies that the observed mass seggregation in these clusters is a dynamical effect. However, larger samples will improve the statistics and give us a better insight in the physical processes leading to the structure and dynamical evolution of clusters.

\section{Acknowledgements}
\acknowledgements
     The authors would like to thank the anonymous referee for very useful and important suggestions to improve the paper. 

This publication makes use of data products
from the Two Micron All Sky Survey, which is a joint project
of the University of Massachusetts and the Infrared Processing
and Analysis Center/California Institute of Technology, funded
by the National Aeronautics and Space Administration and the
National Science Foundation. This research has made use of the WEBDA database, operated at the Institute for Astronomy at the University of Vienna (http://www.univie.ac.at/webda) founded by J.-C.Mermilloid (1988, 1992) 
devoted to observational data on galactic open clusters.) Virtual observatory tools like Aladin and Topcat have been used in the analysis.
This research has been funded by the Department of Science and Technology (DST), India under the Women Scientist Scheme (PH).

\bibliographystyle{spr-mp-nameyear-cnd}
\bibliography{biblio-u1}

\end{document}